# Benthic inputs as predictors of seagrass (*Posidonia oceanica*) fish farm-induced decline

DIAZ-ALMELA[a]* Elena, Núria MARBÀ[a], Elvira ÁLVAREZ[b], Rocío SANTIAGO[a], Marianne HOLMER[c], Antoni GRAU[b], Roberto DANOVARO[d], Marina ARGYROU[e], Ioannis KARAKASSIS[f], and Carlos Manuel DUARTE[a]

a. Interdisciplinary Oceanography Group (GOI). IMEDEA (CSIC-UIB). C/ Miquel Marqués nº 21. 07190, Esporles, Spain. elena.diaz-almela@uib.es ; nuria.marba@uib.es ; rocio.santiago@uib.es ; carlosduarte@imedea.uib.es

b. Department of Fisheries (DGP-CAIB). C/ Foners nº10, 07006, Palma de Mallorca, Spain. ealvarez@dgpesca.caib.es ; agrau@dgpesca.caib.es

c. Institute of Biology, University of Southern Denmark. Campusvej 55, DK-5230 Odense M, Denmark. holmer@biology.sdu.dk

d. Department of Marine Sciences, Polytechnic University of Marche. Via Brecce Bianche, 60131, Ancona, Italy. danovaro@univpm.it

e. Department of Fisheries and Marine Research (DFMR), Ministry of Agriculture, Natural Resources and Environment. 13 Eolou Street, 1416 Nicosia, Cyprus. margyrou@dfmr.moa.gov.cy

f. Department of Biology, University of Crete, P.O. Box 2208, 71409 Heraklion, Crete, Greece. karakassis@biology.uoc.gr

*Corresponding author: elena.diaz-almela@uib.es ,Telephone: +34971611829 ; FAX: +34971611761. C/ Miquel Marqués nº 21. 07190, Esporles, Spain.

# Benthic inputs as predictors of seagrass (*Posidonia oceanica*) fish farm-induced decline

**Abstract**

Fish farms represent a growing source of disturbance to shallow benthic ecosystems like seagrass meadows. Despite some existing insights on the mechanisms underlying decline, efficient tools to quantitatively predict the response of benthic communities to fish farm effluents have not yet been developed. We explored relationships of fish farm organic and nutrient input rates to the sediments with population dynamics of the key seagrass species (*Posidonia oceanica*) in deep meadows growing around four Mediterranean Sea bream and Sea bass fish farms. We performed 2 annual shoot censuses on permanent plots at increasing distance from cages. Before each census we measured sedimentation rates adjacent to the plots using benthic sediment traps. High shoot mortality rates were recorded near the cages, up to 20 times greater than at control sites. Recruitment rates remained similar to undisturbed meadows and could not compensate mortality, leading to rapid seagrass decline within the first 100 meters from cages. Seagrass mortality increased with total ($R^2$= 0.47, $p$< 0.0002), organic matter ($R^2$= 0.36, $p$= 0.001), nitrogen ($R^2$= 0.34, $p$= 0.002) and phosphorus ($R^2$= 0.58, $p$< $3 \cdot 10^{-5}$) sedimentation rates. *P. oceanica* decline accelerated above a phosphorus loading threshold of 50 mg $m^{-2}$ $day^{-1}$. Benthic sedimentation rates seem a powerful predictor of seagrass mortality from fish farming, integrating local hydrodynamics, waste effluents variability and several environmental mechanisms, fuelled by organic inputs and leading to seagrass loss. Coupling direct measurements of benthic sedimentation rates with dynamics of key species is proposed as an efficient way to predict and minimize fish farm impacts to benthic communities.

## Introduction

Marine fish farming has developed rapidly across the world during the last decades and this trend is expected to continue. Around the Mediterranean it has rapidly increased since 1990 and is predicted to grow by 5% annually over the next two decades (UNEP, 2002). In addition, more than half of fish-farm production takes place in coastal waters (UNEP, 2002). Fish cages enhance the input of organic matter and nutrients to the water and nearby sediments, mainly through the release of fish faeces and excess feed pellets (e.g. Wu, 1995; Holmer et al., 2002; in press). Such organic loading immediately affect sediment biogeochemical processes in the vicinity (e.g. La Rosa et al., 2004; Frederiksen et al., 2005), through increasing oxygen consumption (e.g. Holmer et al., 2002) and thereby promoting anaerobic degradation of organic matter (e.g. Holmer et al., 2003a). The depletion of sediment oxygen and excess of reduced toxic products from anaerobic pathways (such as sulphides and ammonium) have an impact on benthic communities (Delgado et al., 1997; Terrados et al., 1999; Karakassis et al., 2000, 2002; Ruiz et al., 2001; La Rosa et al., 2001, 2004; Mirto et al., 2002; Vezzuli et al., 2002; Holmer et al., 2003b).

*Posidonia oceanica,* the dominant and endemic seagrass species in the Mediterranean Sea, extends from 0.3 to 45 meters depth in clear waters, which is also the region preferred for fish farm developments. As other seagrasses, *P. oceanica* is a key species forming meadow communities of high diversity (Templado, 1984), which provide important ecosystem functions (Hemminga and Duarte 2000). Such functions are being jeopardised by the tendency towards a substantial decline of these ecosystems, at rates of about 5% $yr^{-1}$ (Marbà et al., 2005), faster than the 2% $yr^{-1}$ global rate of decline of seagrass ecosystems (Duarte et al., in press).

*P. oceanica* meadows are highly vulnerable to marine aquaculture activities (Holmer et al., 2003b), as reflected by large-scale losses of *P. oceanica* around shallow and sheltered fish farms (e.g. Dimech 2000; Ruiz et al., 2001), which continue even after farming cessation and water quality recovery (Delgado et al., 1999). In an attempt to minimize the impact to the benthos, new farms have been established in recent years at deeper and more exposed sites. This strategy seems to have succeeded for some benthic communities. Alas, this may not be the case for deep *P. oceanica*, whose carbon balance is easily disturbed. Fish farm impacts on *P. oceanica* meadows near their depth limit, however, have not yet been documented.

*P. oceanica,* with its sparse sexual reproduction (e.g. Diaz-Almela et al., 2006), is the slowest-growing seagrass species (Marbà and Duarte, 1998), requiring centuries to (re)colonise coastal areas (e.g. Meinesz and Lefevre, 1984; Duarte 1995; Marbà et al., 2002; Kendrick et al., 2005). Thus, any losses of *P. oceanica* meadows can be considered irreversible at managerial time scales. It is, therefore, essential to develop early indicators of aquaculture-derived impacts to *P. oceanica* meadows, in order to be able to act before irreversible losses occur. Seagrass cover and density have been used in most monitoring programs as indicators of population disturbance (Short and Coles, 2001), but these techniques often detect decline only after significant losses have already occurred, which are then difficult to recover for this slow-growing species. Moreover, previous investigations suggest that discontinuity of farming operations upon observation of losses in *P. oceanica* cover and density are inefficient remedial measures, as losses continue even after fish farm removal (Delgado et al., 1999). This is probably due to the slow recovery of sediment conditions (Delgado et al., 1999; Karakassis et al., 1999). On the other hand, valuable information on the seagrass decline mechanisms, such as whether it operates through increased shoot mortality or reduced recruitment, is lacking in such approaches.

While detailed studies using direct metrics of seagrass dynamics, such as shoot demography, are essential; individual studies can hardly be used to predict the impacts of other farms, as these differ greatly in extent and intensity of impacts on benthic systems (e.g. Ruiz et al., 2001; Karakassis et al., 2002; Crawford et al., 2003). However, a comparative approach across multiple fish farms, yet to be attempted, may lead to the discovery of general relationships predicting the effect of fish farms on seagrass beds. Aquaculture effluents are unanimously considered the main drivers of benthic impacts (e.g. Wu, 1995; Cancemi et al., 2003; Dimech et al., 2000), but few studies have related quantitatively benthic input rates with benthic impacts (Crawford et al., 2003; Holmer et al., 2003a).

In the present effort we examine the impacts of farming activities on shoot population dynamics of a key seagrass (*Posidonia oceanica*) across four deep (16 to 28 m) fish farms around the Mediterranean. We aim to establish a general relationship connecting *P. oceanica* population dynamics with benthic organic loading and nutrient input rates from caged fish farms. This relationship should allow us to predict the impacts of Mediterranean fish farms on *P. oceanica* meadow systems.

## Materials and Methods

We assessed the demography of the seagrass (*Posidonia oceanica*) in deep meadows growing around 4 fish farms, widely distributed along the Mediterranean, from Cyprus to Spain (Fig. 1). The sediments were fine to coarse grained and carbonate-rich (41-46% in Cyprus, > 95% at other sites) and the water depths varied between 16 and 28 m (Table 1). All the farms initiated their activities in the nineties (Table 1); they consisted of 20-24 net cages with an annual production of 260-1150 tones (table 1). The cultured species were gilthead (*Sparus aurata*), sharpsnout sea bream (*Diplodus puntazzo*) and sea bass (*Dicentrarchus labrax*), which were fed with dry pellets. The farms in Cyprus, Italy and Spain were located on open coasts about 1 km from shores, whereas the farm in Greece was located in a strait about 300 m from shore (Fig. 1). In all the sites, the main currents were parallel to the coast, ranging from 8.59 cm $s^{-1}$ (Greece) to more than 20 cm $s^{-1}$ (Italy, Table 1). Further information on fish farm characteristics and environmental conditions around the fish farms is reported in Table 1 and elsewhere (Frederiksen, 2005; Pitta et al., 2006; Holmer et al., in press).

In each site two transects extending from the edge of the meadow closest to the farm to 1000 or 1200 meters away, were established. In Cyprus and Italy both transects extended parallel, in the direction of the main current, while in the Greek and Spanish sites, in order to explore a wider set of conditions, one transect was extended in the direction of the main current while the other was perpendicular to it, towards the coast (Fig. 1).

At each of these transects, we defined three stations: a “disturbed station” was installed in the area vegetated by *P. oceanica* that was closest to the fish farm. This was located, at the time of the study, 5 to 15 m away from the net cages across sites, where sparse plants were found. An “intermediate station”, installed at 20 to 50 meters distance from fish cages, where seagrass beds were denser but not yet comparable to those found further away from the farm. Finally, a “control station”, located at 800-1200 meters distance to fish cages, where no impacts were evident upon visual inspection. This pattern differed at the Cyprus site, since the fish cages were located over deeper bottoms (40 meters depth). Although extremely sparse *P. oceanica* shoots were found close to those cages, the high depth made impracticable to census them. Moreover, *P. oceanica* formed sparse patches until 300 meters away from the cages, towards the shore, forming then a continuous meadow from 20 meters depth upwards. Therefore, at this site, the stations were installed 300 400 and 1000 meters away from fish cages, respectively, in the direction of the main current (Fig. 1).

*Posidonia oceanica* demography

Within each station we installed three permanent plots at the bottom, by SCUBA diving, using metal sticks, ropes and buoys, as explained in Marbà et al. (2005). The size of the triplicate quadrats was adjusted to encompass at least 100 shoots per quadrat (from 0.25 $m^2$ in "control" stations to 7 $m^2$ in Spanish "disturbed" stations). We performed two direct censuses of the shoots present within those permanent plots in each site. Censuses were separated by a period of about one year (from 307 to 386 days, Table 1). During each census, we counted the total number of alive shoots within the plots. As shoot recruitment in *P. oceanica* occurs by apical bifurcation of vertical and horizontal shoots (the latter called apices), we counted the recently bifurcated vertical shoots and all the apices among the total shoot population within the plot and tagged them 2 cm below the meristems with plastic cable ties (10 cm long). Therefore we modified the procedure described in Marbà et al. (2005) to allow the most efficient possible use of the limited bottom time of SCUBA diving at those depths. Tagging allowed us to discriminate the new recruits (unmarked bifurcated shoots and apex bifurcations) in the second census, the new apices (produced by transformation of shoots from vertical to a horizontal growth mode or by the entering of a horizontal apex from outside the plot and, thus, lacking marks) and the total, surviving shoots and apices. We calibrated the counting error by counting 2 plots by independent observers, yielding an estimated error of ± 0.2% and ± 3.5% of the total shoot population for recruits and lost shoots, respectively.

The repeated censuses allowed direct estimates of specific rates ($yr^{-1}$) of shoot mortality and recruitment and net population growth, as well as the expected time to loose half of the shoots at each station.

The specific shoot mortality rate ($M$, in year$^{-1}$, $yr^{-1}$) was calculated as:

$$M = -\frac{(\ln N_{S1}/N_{t0}) \cdot 365}{t_1 - t_0} \qquad (1)$$

Where $N_{t0}$ is the total number of shoots (vertical and horizontal apices) counted in the initial census ($t_0$, days) at each plot, $N_{S1}$ the total number of survivor shoots (vertical and apices) at the second census ($t_1$, days).

The specific shoot recruitment rate ($R$, in $yr^{-1}$) was estimated as:

$$R = \frac{\ln\left((N_{r1} + N_{s1})/N_{s1}\right) \cdot 365}{t_1 - t_0} \qquad (2)$$

Where $N_{r1}$ is the total number of recruited shoots (i.e. bifurcated vertical shoots and apices) observed at $t_1$, and $N_{s1}$ is the number of survivors at $t_1$.

Specific net population growth rates ($\mu$) were estimated as:

$$\mu = R - M \qquad (3)$$

*Sedimentation rates*

We measured benthic sedimentation rates at each station and site by deploying benthic sediment traps next to the plots, 1to 3 times in either June or September (the season of maximum production in the farms), for about 48h periods. The sediment traps were designed after Gacia et al. (1999), and consisted of two replicated arrays situated 20 cm above the bottom, each supporting five 20 ml cylindrical glass centrifugation tubes with an aspect ratio of 5 (16 mm diameter), in order to minimize internal resuspension. The contents of 1-3 tubes were combined and collected on a combusted, pre-weighed Whatman GF/F filter. Dry weight of total sediment deposition was obtained after drying the filters at 60 ºC to constant weight. Dry weight of Organic Matter (OM) deposition was measured through combustion of some of the filters. Total P (TP) was obtained after boiling combusted materials in 1 M HCl for 15 min followed by spectrophotometric determination of phosphate (Koroleff, 1983). We analysed the un-combusted filters for total N contents with an elemental analyzer (Iso-Analytical Ltd. United Kingdom). Further information on these analyses and spatial patterns of fish-farm inputs are shown in Holmer et al., in press. We estimated Total matter, OM, N and P sedimentation rates from these measures according to Blomqvist and Håkanson (1981) and Hargrave and Burns (1979), as described in detail in Gacia et al. (1999).

*Statistical analyses*

Differences in shoot densities among stations within the same locality and transect, as well as between censuses within each station were tested for significance using Student t-tests between pairs of samples. We did least squares regression analysis (Type I), using the SSPS 11.0.4 for MacOsX © package (linear regression procedure with y and/or x data log transformed) between shoot population dynamics parameters and sediment input rates. The regression analysis was performed pooling all the data from all the localities and transects. If there is a strong relationship between benthic sedimentation rates and shoot population dynamics, we would expect to detect a significant and strong correlation of those parameters, despite the local differences in depth, current regime, sediment type, annual production etc.

## Results

The meadows ranged broadly in shoot density among sites, but the areas close to the cages were depleted in shoots relative to distant ones. A general pattern of density reduction towards cages was observed across sites except in Cyprus, where the stations closest to the cages were situated 300 meters away. A sharp shoot decline was recorded next to the cages in Sounion (Greece) and Porto Palo (Italy) between censuses. At these sites, the shoot densities at impacted stations (15 and 5 m from cages respectively) decreased from 102 to 14 shoots $m^{-2}$ (90% of reduction) and 128 to 22 shoots $m^{-2}$ (81% of reduction) in 353 and 307 days respectively (Fig. 2). At El Campello (Spain), the deepest location, there was a modest decline in 312 days, which was only significant at the impacted (10 m distance to the cages, $p= 0.03$) and intermediate (40 m to the cages, $p= 0.04$) stations of the transect perpendicular to the coast and to the main current (Fig. 2). In Cyprus, shoot density only declined significantly in one of the impacted stations, and this decline was much lower than at the other sites (Fig.2). Shoot density in control stations was much more stable between censuses. Significant reductions in shoot numbers were only observed in one control station in Greece ($p < 0.05$), and shoot density slightly increased at the control station in one of the transects at the Spanish farm ($p= 0.003$, Fig. 2). Intermediate stations exhibited intermediate behaviours except in Greece, where the plots situated perpendicular to the current, experienced a large (30%) increase in shoot density.

Shoot density decline was driven by high shoot mortality rates, reaching 4.19 ± 1.77 (SE) $yr^{-1}$ at one of the impacted stations in Italy (Fig. 3). Shoot mortality rates were, on average, 7.5 (Spain) to 19.4 times (Greece) higher at the impacted stations near fish cages than at control stations (Fig. 3). Relative shoot recruitment rates ranged between 0.01 and 0.31 $yr^{-1}$ among sites and stations.

No clear spatial patterns were detected in shoot recruitment, which variability increased near the cages (Fig. 3). The observed shoot recruitment near fish cages was between 3 and 300 times lower than mortality and therefore could not compensate the losses. This leaded to high net decline rates of the shoot population in most impacted and intermediate stations (Fig. 3).

Shoot mortality and net population decline rates (and consequently also shoot half life) decreased as a power law of the distance to cages across sites ($R^2$= 0.63, $p$< $10^{-6}$; $R^2$= 0.57, p< $2 \cdot 10^{-4}$ respectively; Fig. 4a, 4c, Table 2).

Total, Organic matter, N and P benthic benthic sedimentation rates exponentially declined with distance to cages (Fig. 5).

Specific shoot mortality rates increased exponentially with phosphorus sedimentation rate ($R^2$=0.57; $p$< 0.001, Fig. 6, Table 2). The specific mortality rate was also significantly correlated, although less strongly, with the input of total ($R^2$= 0.47 ; $p$< 0.001) and organic ($R^2$= 0.36; $p$= 0.001) matter and nitrogen ($R^2$= 0.34 ; $p$= 0.002, Fig. 6, Table 2). Shoot recruitment rates decreased exponentially with increasing Organic Matter inputs, but the correlation was low ($R^2$= 0.19, p< 0.03; Table 2). The correlation with N inputs was marginally significant (P= 0.05).

**Discussion**

The results presented here clearly demonstrate a dramatic impact of fish farm wastes on deep *P. oceanica* seagrass meadows. In less than 1 year, the extension of bare sediment and vegetated area with reduced shoot density increased outwards from the cages. This pattern of decline was particularly evident at the Italian fish farm, the largest one and with the highest production, where shoot density at the intermediate station (40 m from cages) reached the levels recorded at impacted stations (5 m) after 1 year. A similar, strong regression of *P. oceanica* meadow is reported by Delgado et al. (1999) around a small, shallow sheltered fish farm.

Shoot mortality and decline rates rapidly decreased with distance from the farms. The regression suggested a reduction by half of those parameters at 80 meters distance, when compared to the rates beneath the cages. Consequently, the seagrass shoot half-life significantly increased within the first 100 meters from the cages, indicating that, beyond this distance, decline is much slower. Nevertheless as the curve of mortality did not reach the global mean+SD recruitment rates (0.13+0.10 year$^{-1}$) until 400 meters from the cages, the balanced seagrass growth ($R = M$) may be only achieved beyond this distance. This finding is consistent with the observation by Marbà et al. (2006) that the concurrent rhizome vertical growth in the same sites is reduced by half after the farm onset at distances as high as 300 m from fish cages and that, in the largest farm, the shoot growth reduction is still significant 1000 meters away (Marbà et al., 2006).

On the other hand, meadow decline was very fast near the cages. The relationship between shoot half life and distance from the cages predicts shoot density to decline by half in about 3 to 26 months within the first 15 meters from the cages, and in 1 to 6 years, on average, within the first 50 meters from the cages. The regression equation describing the increase in net population growth rates with distance from the fish cages predicts that meadows would be lost (i.e. density reduced by > 90 %) after 5 to 11 years and 11 to 32 years, on average, within the first 15 and 50 meters from cages respectively, at the studied farms. Such predictions are based on the declining rates registered during only one year and therefore they do not take into account the possible temporal variability of the declining rate. For instance decline could accelerate with the reduction of meadow cover, as suggested by Duarte et al. (2002). Nevertheless, the population dynamics approach allowed us to predict the magnitude and velocity of future decline and to give some insight on the mechanism i.e. high shoot mortality not coped by recruitment.

There was also substantial variability in decline responses to distance to cages among sites and stations, as reflected in the residuals of the regression lines describing the general relationships. This variability likely reflects patchiness in the distribution of the impacts, dependent for instance on the local variability of current patterns.

Significant net decline was still recorded at the control station closest to the coast in Sounion (Greece), and, although not significant, relatively high declines occurred in the control station of the largest fish farm (Italy). However, these declines cannot be exclusively attributed to fish farm influence, as there were other potential sources of impact (e.g. a sewage outfall approximately 2 miles away from the Italian fish cages and 1 mile from the control stations. The decline rates recorded at control stations are comparable to rates documented for other *P. oceanica* meadows without fish farm influences across the Mediterranean (Marbà et al., 2005).

The extension of seagrass die-off and density reductions observed here are similar to those documented around shallow and sheltered small Mediterranean fish farms producing less than 100 $T\ yr^{-1}$ of fish (Delgado et al., 1997, 1999; Pergent et al., 1999; Dimech et al., 2000) and around deep farms of similar production (200 $T\ yr^{-1}$, Pergent et al., 1999). The extension of the impact of deep farms would be reduced if compared with shallow and sheltered larger fish farms. For example, the linear extension of seagrass affected by a fish farm producing 700-800 $T\ yr^{-1}$ of sea bream and sea bass in a shallow bay during 8 years is more than 2 times longer (reaching more than 200 meters away from fish cages, thus the area affected would be ca 4 to 9 times greater) (Ruiz et al., 2001) than observed in the deep farms from this study. As previously hypothesized (e.g. Maldonado et al., 2005) this could be explained by the increased dilution of the waste products and, consequently, lower inputs to the sediments in deep sites. Nevertheless, the extension of fish farm impacts on deep meadows of this key species were not limited to the area beneath the cages, a case of figure that differs from  other benthic s, as macro-invertebrates (Maldonado et al., 2005).

Shoot mortality and net population decline increased with increasing sedimentation rates. Total, organic matter and nutrient inputs directly measured on the meadows proved to be useful predictors of seagrass decline rates. Fish farms release significant amounts of waste products, as the feeding efficiency is usually low, with feed conversion ratios (FCR) ranging from 1.1 in efficient cultures of Salmon (Nordgarden et al., 2003) to 6.5 in cultures of aerolated grouper (Leung et al., 1999). FCR in the sea bream cages from the study had intermediate values, ranging from 1.6 to 2.4 (Holmer et al., in press). Much of the materials delivered (feed pellets, faeces and excretion products) reach the sediments, where nutrient loadings have been shown to increase with FCR, (Islam, 2005), thereby affecting benthic communities (Holmer et al., 2002, 2003a). Nevertheless there was a substantial variability in decline responses to benthic loading among sites and transects, as reflected by the residuals of mortality around predictions from benthic inputs. This suggests that we cannot neglect the importance of other local factors on the response of seagrass meadows to fish farms.

The power of sedimentation inputs to predict seagrass demography would derive from the fact that the extent and shape of the fish farm benthic load depends on distance to fish cages (Holmer et al., in press; this work), but also integrates the local effects of depth, fish farm effluent type or quantity and hydrodynamics. Moreover, several mechanisms that have been suggested or shown to have deleterious effects on seagrass (like sediment organic enrichment, Delgado et al., 1999; Cancemi et al., 2000, 2003; herbivore pressure, Delgado et al., 1997; Ruiz et al., 2001; sediment anoxia, Greve et al., 2003; or pore water sulphide, Halun et al., 2002) have been recently shown to be fuelled by fish farm inputs, at least in the farms at this study. Holmer et al. (in press) showed that sediment organic matter and phosphorus contents increase with fish farm loadings in the meadows studied here. Frederiksen (2005) showed that sulphate reduction rates and acid-volatile sulphides in the sediment, as well as the depth of the sulphide front are correlated with organic input rates. Moreover this author observed a significant increase of plant sulphur content with fish farm inputs and correlations between plant sulphur content and mortality rates in the Greek and Italian sites. Thus sedimentary inputs of organic matter and nutrients integrate multiple cooperative impacts on seagrass dynamics through the different mechanisms associated with these inputs, thereby allowing the prediction of seagrass decline. Among the various benthic loadings that have been investigated, phosphorus input rate seems the best parameter to predict seagrass decline, as it explains a large fraction of the variance in shoot decline rates (59%) across sites.

Closer examination of the relationship between seagrass mortality and sedimentary inputs described here (Fig. 6) suggests the existence of thresholds of nutrient inputs above which seagrass decline is accelerated. In *P. oceanica*, mortality accelerates and recruitment declines above input rates of 50 mg P $m^{-2}$ $day^{-1}$ or 1.5 g organic matter $m^{-2}$ $day^{-1}$. Such thresholds of inputs for balanced meadow dynamics (which integrate background and fish farm inputs) may provide a powerful tool to set targets to regulate the location and size of new fish farms in the Mediterranean, and to manage existing ones in a sustainable way.

The examination of the shoot population dynamics of a seagrass within permanent plots (which minimize error) and the deployment of benthic sediment traps could represent, as demonstrated here, an efficient strategy to monitor slow-growing seagrass meadows near fish farms, allowing early detection of impacts which would enable remedial actions preventing further losses.

This study was restricted to sites with sea bream and sea bass fish cages and with an endemic and vulnerable benthic community (*Posidonia oceanica* meadows). However, the findings from studies in other kinds of aquaculture exploitation are consistent with the hypothesis of a strong relationship between benthic fish farm inputs and impact. For example the little or no effect on benthic ecosystems (including seagrasses) under shellfish cages in Tasmania was related by Crawford et al. (2003) to the observation that these cultures of filter feeding organisms did not increase significantly benthic sedimentation with respect to background. The hypothesis is also consistent with the relatively higher impact of similar farming productions on shallow sheltered sites (Ruiz et al., 2001), where fish farm inputs to the sediments are expected to be high due to the lower dispersion of effluents.

Therefore we believe that the approach advanced here, directly linking organic and nutrient (particularly phosphorus) input rates to key-benthic species, could be extended to other benthic  communities and farms (or even pollution sources), as an efficient and simple way to predict benthic ecosystem impacts, allowing to accurately define effluent thresholds to sustainable activities.

**Acknowledgements**

This research was funded by projects MedVeg (Q5RS-2001-02456 of FP5) and THRESHOLDS (contract 003933-2 of FP 6) of the European Union. We are grateful to A. Petrou, Y. Nicolaou, S. Savvas, K. Tochtarides, C. Basilakoulos, A. Grafas, T. Proutzos, G. Skandamis, M. Salomidi, J. Glampedakis, M. Tsapakis, S. Mirto, G. M. Luna, F. M. Perrone, C. Corinaldesi, M. Pisconti, L. Bongiorni, C. Vasapollo, R. Martínez, F. Lázaro, A. Rabito, J. M. Ruiz and O. Invers, for assistance in the field.

Table headings:

Table 1 Characteristics of the Mediterranean fish farm studied. FCR = Food Conversion Ratio (the ratio of food supplied to fish production by weight).

Table 2. Fitted least squares regression equations ($y= e^{a} \square e^{bx}$ (or $y= e^{a}\square x^{b}$ for coefficients marked in italics) describing the relationships between P. oceanica shoot mortality, shoot recruitment and net declining rates (y variables; year-1) with distance from the fish farm, total, organic matter, nitrogen and phosphorus input rate (X variables). Coefficients a and b are given with standard errors.

Table 1.

| | Cyprus | Greece | Italy | Spain |
|---|---|---|---|---|
| Location | Limassol | Sounion | Porto Palo | El Campello |
| Fish farm initiated in: | 1992 | 1996 | 1993-94 | 1995 |
| Annual prod. (Tm) | 150 (300 since 2001) | 400 | 1150 | 260 |
| FCR* | 2.2 | 1.60 | 2.39 | 2.00 |
| Mean current speed (cm $s^{-1}$) | 10-15 | 8.9 | >20 (40% of time) | 9.8 |
| Days between censuses | 386 | 353 | 307 | 312 |
| Depth (m) | 19-20 (fish cages: 39 m) | 14-16 | 21-22 | 26-28 |

Table 2.

| | Mortality rate ($year^{-1}$) | Recruitment rate ($year^{-1}$) | Net decline rate ($year^{-1}$) |
|---|---|---|---|
| 1. Total sed. rate | $R^2$= 0.47 $p$< 0.0002 (n= 24) | *ns* | $R^2$= 0.45 $p$< 0.002 (n = 18) |
| (g(DW) $m^{-2}$ $day^{-1}$) | b = 0.28 ± 0.06 a = -2.80± 0.42 | | b = 0.43 ± 0.11 a = -4.34 ± 0.81 |
| 2. OM sed. rate | $R^2$= 0.36 $p$= 0.001 (n = 24) | *$R^2$= 0.19 p= 0.02 (n= 24)* | $R^2$= 0.28 $p$< 0.02 (n = 18) |
| (g(DW) $m^{-2}$ $day^{-1}$) | b = 0.54 ± 0.14 a = -2.06 ± 0.34 | *b = -0.54 ± 0.22 a = -2.01± 0.19* | b = 0.67 ± 0.24 a = -2.79 ± 0.65 |
| 3. N sed. rate (g(DW) | $R^2$= 0.34 $p$< 0.002 (n = 24) | *ns ($R^2$= 0.11 p= 0.06 (n= 24))* | *$R^2$= 0.35 p< 0.007 (n = 18)* |
| $m^{-2}$ $day^{-1}$) | b = 16.25 ± 4.49 a = -1.82 ± 0.30 | | *b = 1.22 ± 0.39 a = 0* |
| 4. P sed. rate | $R^2$= 0.58 $p$< $2 \cdot 10^{-5}$ (n = 23) | *ns* | *$R^2$= 0.59 p< 0.0002 (n = 18)* |
| (mg(DW) $m^{-2}$ $day^{-1}$) | b = 21.01 ± 3.78 a = -1.65 ± 0.20 | | *b = 0.97 ± 0.19 a = 2.53 ± 0.79* |
| 6. Distance (m) | *$R^2$= 0.63 p< $3 \cdot 10^{-6}$ (n=24)* | *ns* | *$R^2$= 0.57 p< 0.0002 (n = 18)* |
| | *b = -0.47± 0.08 a = 1.23 ± 0.38* | | *b = -0.63 ± 0.13 a = 1.47 ± 0.62* |

**Figure legends**

Fig. 1: Locations of the fish farm sites analysed in this study. Filled circle: El Campello (Spain), filled square: Porto Palo (Italy), open circle: Sounion (Greece), open square: Amathous (Cyprus).

Fig. 2: Average (± standard error) percentage of variation in *P. oceanica* shoot density at impacted (lined bars), intermediate (grey bars) and control stations (white bars) from Amathous (Cyprus), Porto Palo (Italy), Sounion (Greece) and El Campello (Spain), measured using repeated censuses within the permanent plots. Within each station, the bar on the left correspond to station from transect 1, and the right one to station from transect 2. Significance levels of density change are indicated (*: $p<0.05$, **: $p<0.01$), as well as the distances of the stations from fish cages.

Fig. 3: Distribution of *P. oceanica* shoot demographic parameters (*M*, specific shoot mortality, *R*, recruitment and $\mu$, net population growth rates) with distance from the fish cages for the various fish farms examined. The boxes indicate the range of shoot mortality, recruitment and net population growth within the station, with upper, middle and lower horizontal lines of the boxes indicating the values recorded in each of the three station plots. Lined bars correspond to impacted stations, grey bars to intermediate stations and white bars to control stations. Within each station, the first bar corresponds to transect 1 and the second bar to transect 2.

Fig. 4: Variability of *P. oceanica* shoot demographic dynamics (specific shoot mortality, *M*; recruitment, *R* and net population decline rates, $-\mu$, year$^{-1}$) with distance ( m) from the fish cages. Symbols correspond to the sites as represented in Figure 1.

Fig. 5: Benthic sedimentation rates (g (dry weight) $m^2$ $day^{-1}$) in relation to distance to fish cages (meters). OM: Organic matter, N: Nitrogen, P: Phosphorus. Lines show the fitted exponential regression lines, when significant: Total ($R^2$=0.36, p<0.0013; b= $-7.4 \cdot 10^{-4} \pm 2.0 \cdot 10^{-4}$, a=7.59 (6.74; 8.56), Nitrogen ($R^2$=0.24, p<0.01; b= -0.002±0.009, a=0.06 (0.04; 0.08) and Phosphorus ($R^2$=0.51, $p<10^{-4}$; b= -0.0022±0.0004, a=0.04 (0.03; 0.05). Data symbols correspond to the sites as represented in Fig. 1.

Fig. 6: The relationship between *P. oceanica* specific mortality (*M*, $year^{-1}$) and sedimentation rates of total, organic matter (OM), nitrogen (N) and phosphorus (P). Lines show the fitted regression lines (Table 2). Data symbols correspond to the sites as represented in Fig. 1.

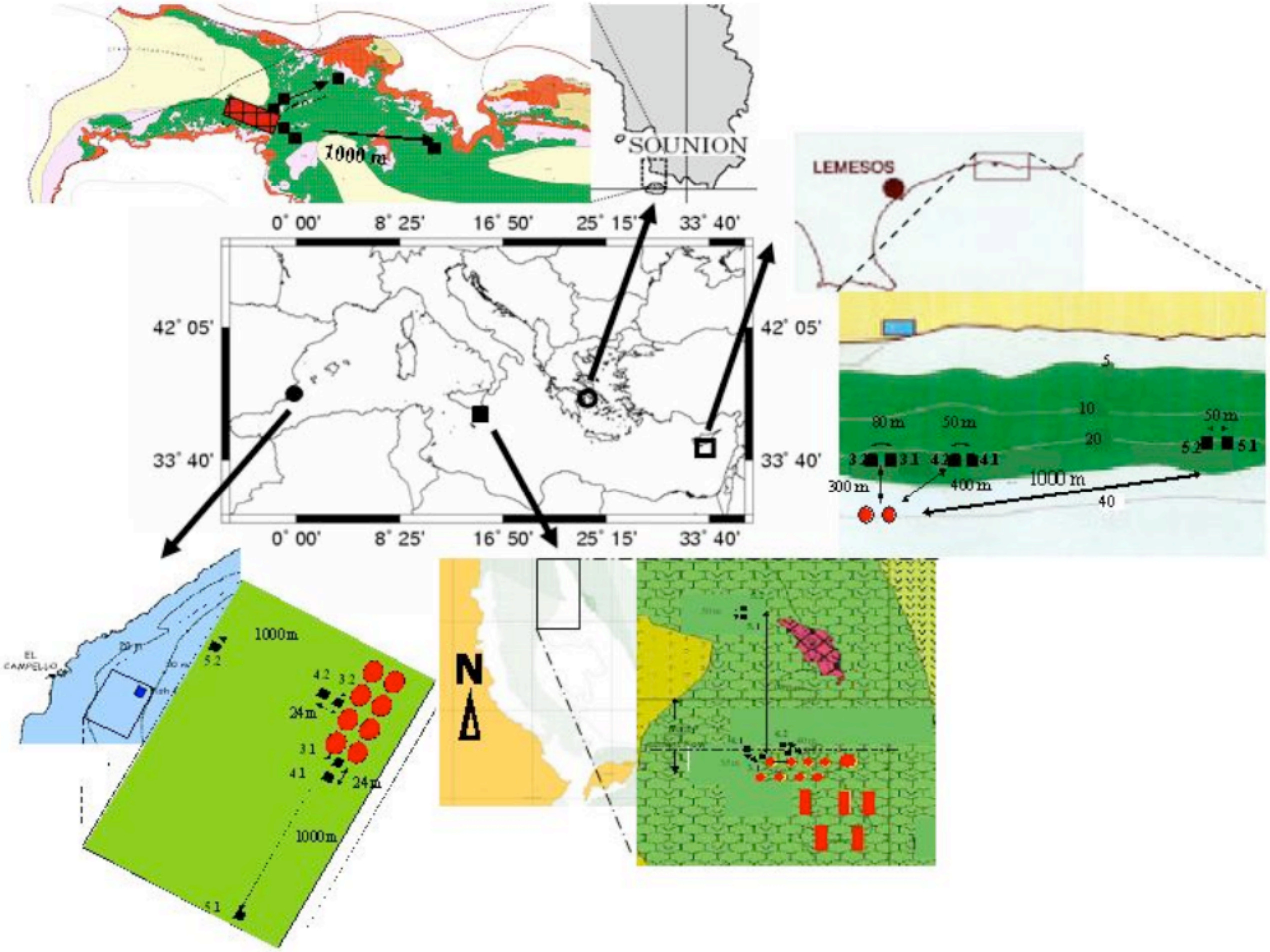

1 Fig. 1

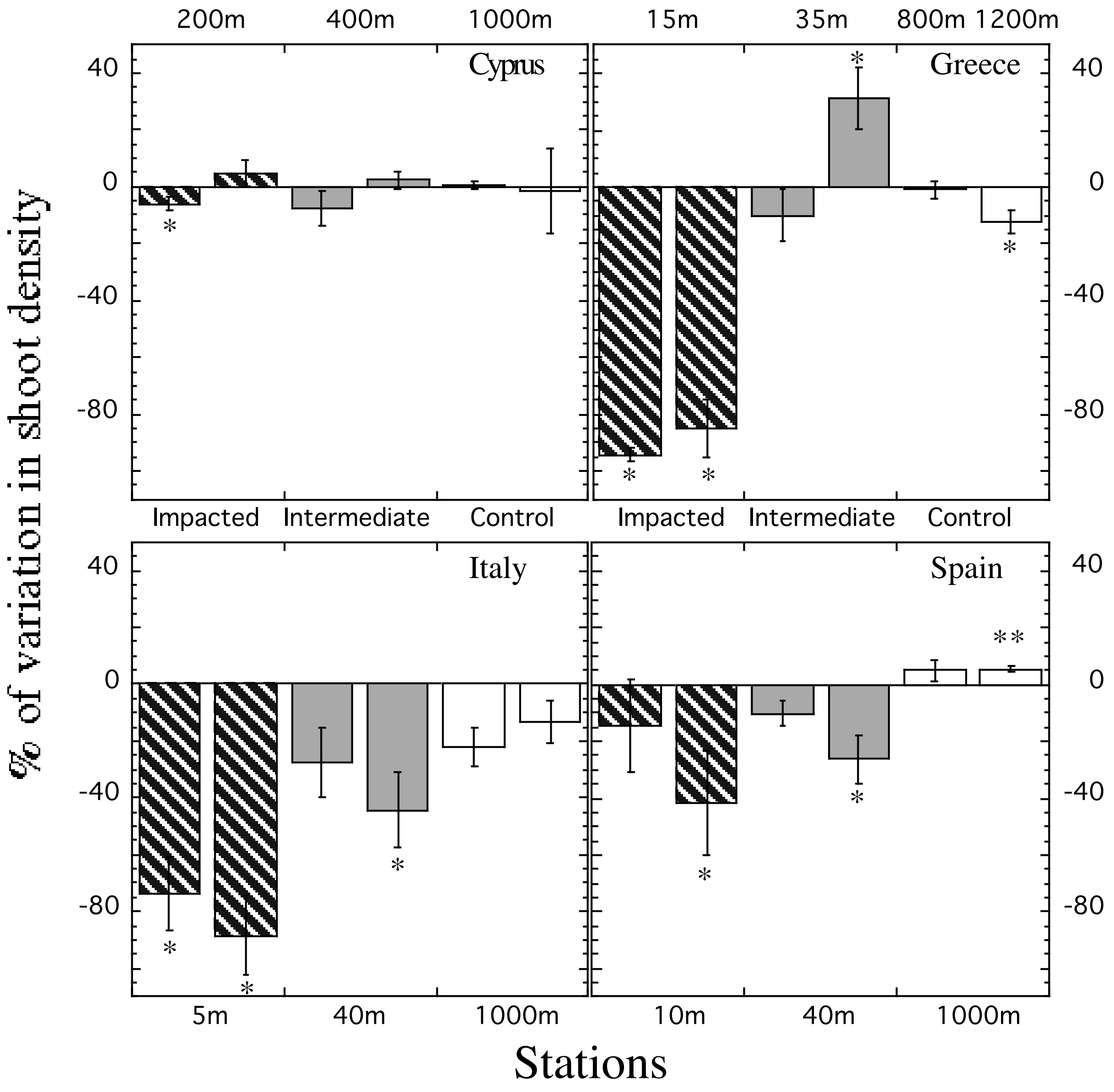

Fig. 2

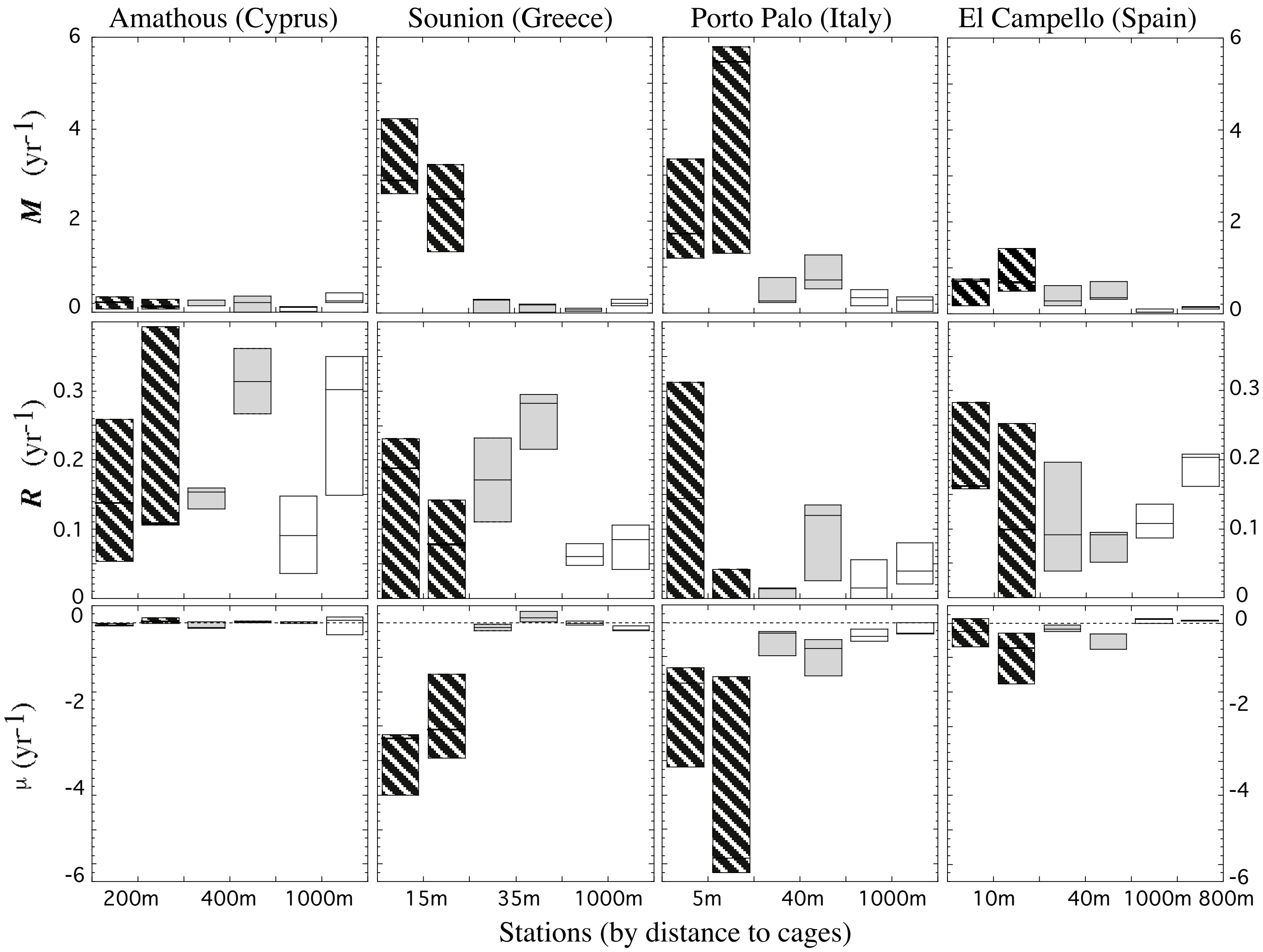

Fig. 3

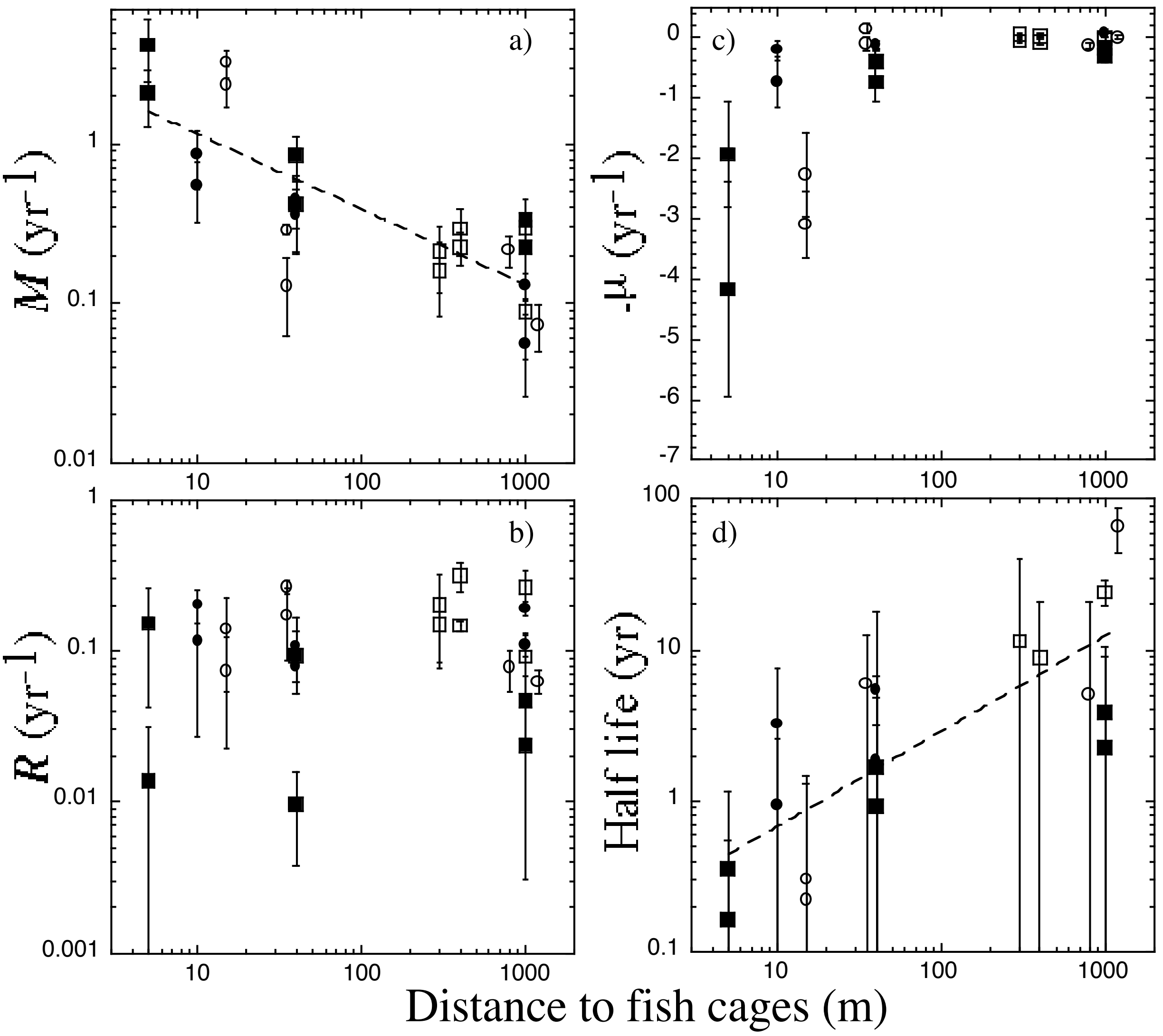

Fig. 4

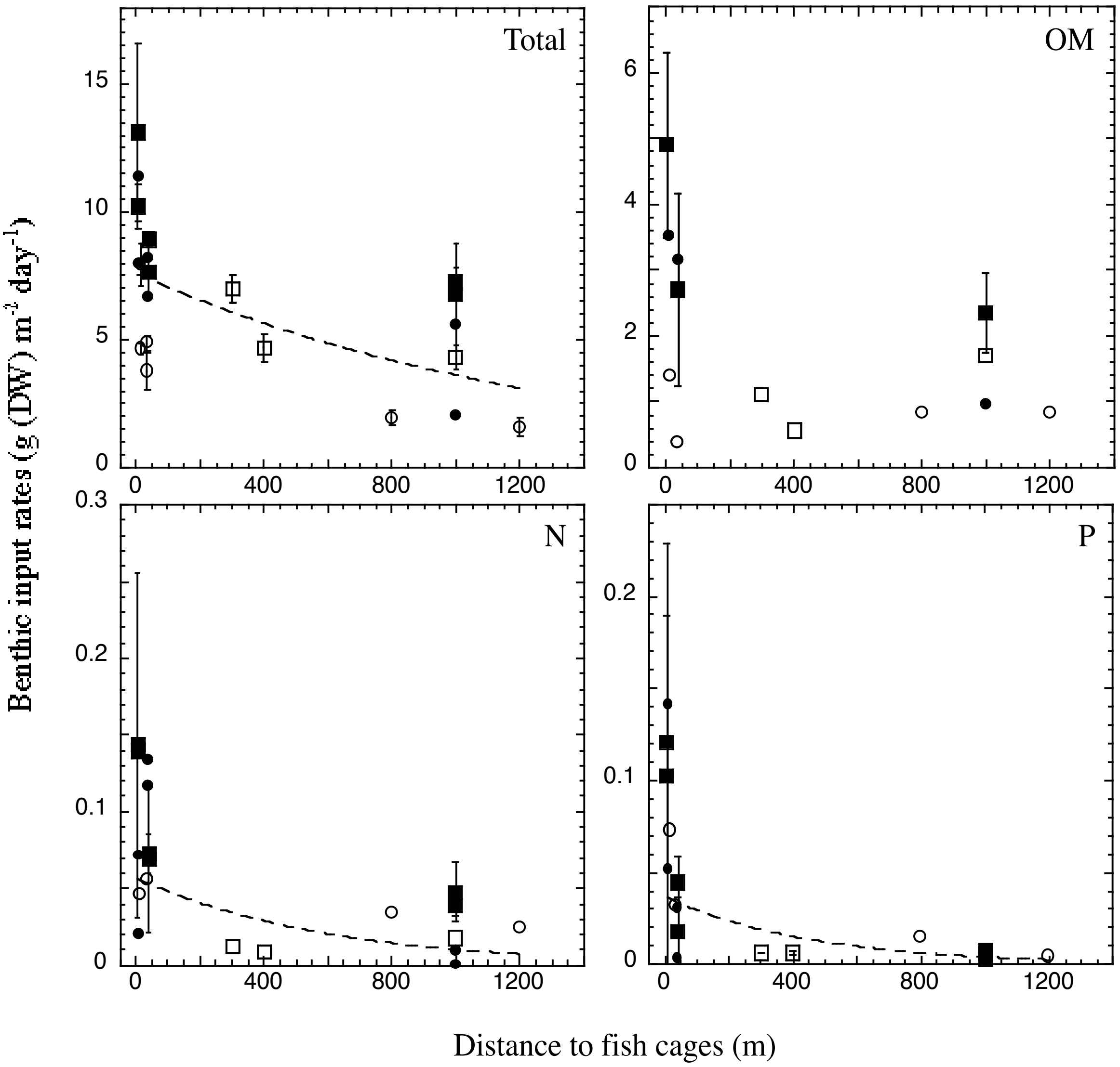

1

2 Fig. 5

3

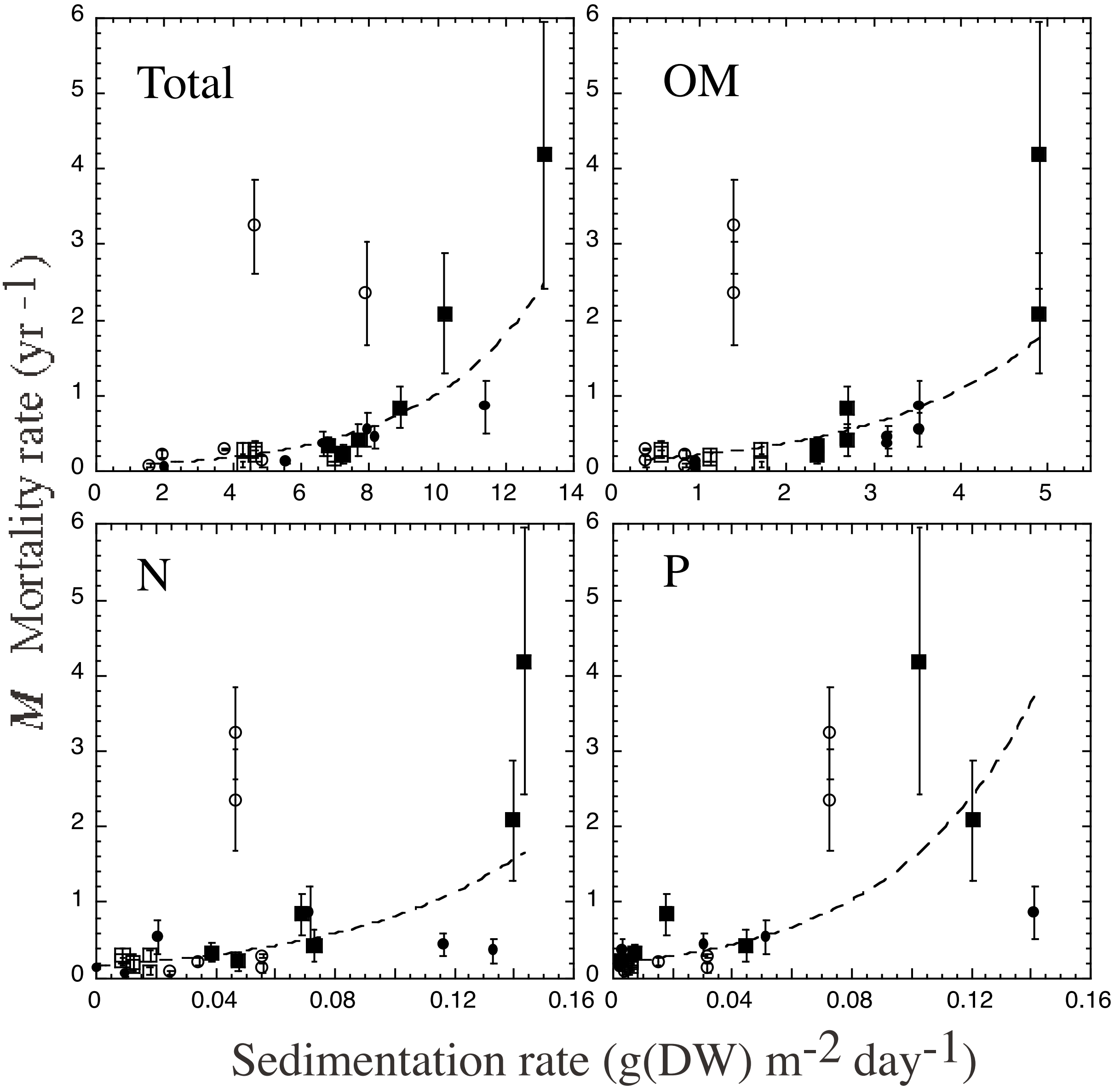

Total
OM
N
P
M Mortality rate (yr -1)
Sedimentation rate (g(DW) m-2 day-1)

1

2 Fig. 6